\documentclass[aps,pra,superscriptaddress,amsmath,amsfonts,amssymb,floatfix,nofootinbib]{revtex4}

\usepackage{enumerate}
\usepackage{mathtools}
\usepackage{amsmath}
\usepackage{graphicx}
\usepackage{epsfig}
\usepackage{sidecap}
\usepackage{hyperref,stackrel}
\usepackage[normalem]{ulem}
\usepackage{color}
\usepackage{enumerate}
\usepackage{bm}
\usepackage[utf8]{inputenc}
\usepackage{amsmath}
\usepackage{amsfonts}
\usepackage{amssymb}
\usepackage{subfig}
\usepackage{graphicx}
\usepackage{enumitem}
\usepackage [autostyle, english = american]{csquotes}
\MakeOuterQuote{"}
\usepackage{br
aket}
\usepackage{caption}
\begin{document}

\title{Hide and seek with quantum resources: New and modified protocols for quantum steganography}
\author{Rohan Joshi} \thanks{E-mail:
joshirohan043@gmail.com}
\affiliation{Delhi Technological University, Shahbad Daulatpur, Main Bawana Road, Delhi-110042, India}
\author{Akhil Gupta} \thanks{E-mail:
guptaakhil.dtu@gmail.com}
\affiliation{Delhi Technological University, Shahbad Daulatpur, Main Bawana Road, Delhi-110042, India}
\author{Kishore Thapliyal} \thanks{E-mail:
tkishore36@gmail.com}
\affiliation{Joint Laboratory of Optics of Palacky University and Institute of Physics of CAS,
Faculty of Science, Palacky University, 17. listopadu 12, 771 46 Olomouc, Czech Republic}

\author{R Srikanth} \thanks{E-mail:
srik@poornaprajna.org}
\affiliation{Theoretical Sciences Division, Poornaprajna Institute of Scientific Research, Bidalur Bengaluru-562164, India}

\author{Anirban  Pathak} \thanks{E-mail:
anirban.pathak@gmail.com}
\affiliation{Jaypee Institute of Information Technology, A-10, Sector-62, Noida, UP-201309, India}

\begin{abstract}
Steganography  is the science of hiding and communicating a secret message by embedding it in an innocent looking text such that the eavesdropper is unaware of its existence. Previously, attempts were made to establish steganography using quantum key distribution (QKD). Recently, it has been shown that such protocols are vulnerable to a certain steganalysis attack that can detect the presence of the hidden message and suppress the entire communication. In this work, we elaborate on the vulnerabilities of the original protocol which make it insecure against this detection attack. Further, we propose a novel steganography protocol using discrete modulation continuous variable QKD that eliminates the threat of this detection-based attack. Deriving from the properties of our protocol, we also propose modifications in the original protocol to dispose of its vulnerabilities and make it insusceptible to steganalysis.
\end{abstract}

\maketitle

\section{Introduction}

Steganography is one of the most fascinating aspects of secure communication. The word "steganography" is derived from the Greek words $steganos$, meaning `covered', and $graphia$, meaning `writing'. Thus, it literally means "covered writing" and  refers to the art of hiding a secret message ($stegotext$) behind an innocuous looking text ($covertext$) in such a way that it can only be detected and deciphered by the intended receiver. It is different from cryptography, where the idea that secret messages are being exchanged is openly known. By contrast, in steganography we aim to hide the existence of the secret message. Steganography has been widely used throughout human history, and examples of the historical use of steganography can be found in abundance in history and also in the animal kingdom (for a short but interesting history of steganography see Ref. \cite{kahn1996history}). Specifically,  in ancient Greece, the messages were marked on shaven heads of trusted messengers who were then sent on their way once the hair had regrown. 

One can point out  several examples of situations where steganography will be of practical use. For example, consider that Bob has decided to give Eve a surprise gift on their anniversary party, despite Eve's challenge that she would find out the contents of the gift earlier. In order to succeed at his attempt, Bob takes the help of Alice, Eve's sister. His only method of communicating his plans to Alice is to send an encrypted invitation. It is not enough that the message is encrypted since it may make Eve and other guests, who are unaware of the plan, suspicious. What Bob needs to save the day is steganography. 

Interesting properties of steganography and its applications in providing security and privacy to internet has drawn considerable attention of the community interested in secure communication (see \cite{cox2007digital, johnson1998exploring, Fridrich2009Nov} and references therein). In fact,  classical steganography has been developed extensively in the later part of 20th century and in the beginning of the present century   \cite{johnson1998exploring, anderson1998limits, sallee2003model, provos2003hide, cachin1998information, Fridrich2009Nov}. In these works, steganography was studied in view of different prospective applications ranging from digital image processing to internet security.

In a different line of research, a protocol for quantum key distribution (QKD) was proposed in 1984 by Bennett and Brassard \cite{BB84}, now known as BB84 protocol, showing that unconditional  security of information can be obtained in the quantum world. This paved the way for several other protocols of QKD (see \cite{shenoy2017quantum} and references therein). This evokes the natural question: Can the advantage of quantum cryptography be extended to design secure protocols of quantum steganography? Addressing this question, in 2002   Gea-Banaloche \cite{Gea-Banacloche2002Sep} proposed the first protocol of quantum steganography. This pioneering work of Gea-Banaloche has been followed by a number of studies on quantum steganography \cite{shaw2011quantum, qu2010novel, jiang2016lsb, mihara2015quantum, luo2019efficient, csahin2018novel, qu2019matrix, el2012quantum, mar}. In these schemes, different strategies involving quantum resources have been used to conceal the stegotext. For example, in  \cite{shaw2011quantum}, the stegotext was concealed by giving it the appearance of channel noise in a codeword of a quantum error-correcting code; whereas in \cite{qu2010novel}, the ping-pong protocol for quantum secure direct communication and entanglement swapping were used to design a scheme of quantum steganography. Further, preshared entanglement  and GHZ states were used as quantum resources in \cite{mihara2015quantum} and in \cite{el2012quantum}, respectively.  \\
All the protocols for quantum steganography proposed in the above mentioned studies and the references therein are expected to fulfill the following requirements:
\begin{enumerate}
    \item \textbf{Communication:} The transmitting party is able to communicate classical or quantum information to the receiving party successfully.
    \item \textbf{Secrecy:} The stegotext is completely concealed such that the eavesdropper or person in authority should be unable to detect its presence.
\end{enumerate}
In addition, the requirement of \textbf{security} can be imposed to ensure that a third party cannot read the stegotext even if its presence is detected. Since steganography focuses only on hiding the fact that a secret message is being transmitted, it is not necessary to encrypt the message, that is why security is a separate criterion.  This criterion is often fulfilled through the use of quantum cryptography. In this regard, the distinction between quantum steganography and quantum cryptography can be further emphasized by stating that while the former requires all three requirements to be satisfied, the latter requires only security as the necessary and sufficient condition. Interestingly, it was shown by Martin \cite{mar} that a quantum steganographic protocol can be integrated within a cryptographic protocol to communicate a hidden classical bit successfully. Hereafter, this protocol will be referred to as Martin's quantum steganography (MQS) protocol. Further,  this protocol may be viewed as a variant of BB84 protocol for QKD \cite{BB84} with a steganographic channel. In what follows we will give specific attention to this scheme. 

It would seem that if Alice and Bob are employing QKD, then they could simply employ QKD to send a secret bit, rather than use steganography. The motivation for the latter arises in the situation where Alice and Bob are prohibited by cost considerations to use intermediate-security QKD equipment, e.g., Noisy Intermediate Scale Quantum (NISQ) tools rather than fully device-independent ones. Thus, with sufficient resources, Eve can gain information about a good fraction of their messages by performing a conventional QKD cryptanalysis. Thus, to transmit top secret bits, they may resort to steganography. 

A steganalysis of MQS protocol has been performed by Qu et al. \cite{stega}, who has reported certain vulnerabilities of MQS protocol and proposed a notion of steganalysis using coherent measures to detect the presence of a hidden channel in open channel. Steganalysis relies on the principle that classical steganography changes the probability distribution of the quantum states. The deviation of the detected probability distribution from the theoretical distribution can be analyzed by quantum state discrimination to achieve effective detection of steganographic communication. The attractive idea about MQS protocol is that in terms of practical implementation, it can be realized with only the elements that make up QKD. Here, we aim to build on the idea of basing steganography on QKD, but free of the vulnerabilities mentioned above.

The rest of the paper is structured as follows. In Section \ref{sec:MQS}, we briefly describe MQS protocol \cite{mar}  and its weaknesses. In Section \ref{sec:steganalysis}, the steganalysis of MQS protocol reported in the existing literature is briefly reviewed to elaborate on its vulnerabilities and the need for a new protocol for quantum steganography free from the weaknesses of MQS protocol. A new protocol for quantum steganography is proposed and analyzed in  Section \ref{sec:new-protocol}. The protocol uses reverse communication for a class of discrete modulation continuous variable-QKD (CV-QKD) protocols which may be realized using coherent states  \cite{Namiki_2003,PhysRevA.74.032302} or other quantum states as quantum resource. The paper is  concluded in Section \ref{sec:conclusion} with a short discussion on the use of reverse communication in circumventing vulnerabilities of  MQS protocol. 

\section{Review of MQS protocol\label{sec:MQS}}
Here, we aim to briefly review the MQS  protocol proposed by Martin and discuss its security. In his protocol, Martin used BB84 QKD protocol for covert communication. Alice and Bob are two parties who wish to establish successful steganographic communication using their QKD channel. The steps of MQS protocol are as follows:\\
\\ 
\textbf{MQS1:} Alice prepares a  random string of $4m$ qubits\footnote{In the original protocol, Martin used $4m$ qubits, but it would have been more practical to use $4(m + \delta)$ qubits to take care of the fluctuations and to ensure that with high proability $2m$ qubits are obtained in Step MQS5.}, where the qubits are prepared randomly in $\{\ket{0}, \ket{1}\}$ or $\{\ket{+}, \ket{-}\}$ basis.\\
\textbf{MQS2:} Alice sends the string to Bob.\\ 
\textbf{MQS3:} After receiving the qubits, Bob measures them randomly in $\{\ket{0},\ket{1}\}$ or $\{\ket{+},\ket{-}\}$  basis.\\ 
\textbf{MQS4:} Alice announces the basis in which she originally encoded her bits.\\
\textbf{MQS5:} Bob announces the positions of bits in which his measurement basis coincides with the encoding basis and keeps the corresponding bits. The number of remaining bits is about ~$2m$. \\
\textbf{MQS6:} Alice decides her stego bit from the remaining $2m$ bits. The value of this bit is the information that Alice wants to send to Bob. She initialises the check-bit method whereby she randomly announces $m-1$ check bits out of the remaining $2m-1$ bits. The $m^{\textrm{th}}$ check bit is chosen such that it lies in a pre-decided spatial relation to the stego bit. For example, the stego bit can lie to the right of the $m^{\textrm{th}}$ check bit with a pre-decided displacement $d$ that is calculated using the key generated in the previous run of QKD.\\
\textbf{MQS7:} Alice and Bob compare the values of their check bits. If the error percentage is more than a certain threshold, then they abort the protocol.\\ 
\textbf{MQS8:} Alice and Bob perform classical post-processing to distill a shorter bit string from remaining $m$ bits.

In this manner, a single stego bit can be communicated from Alice to Bob in one QKD run. Initially, Alice and Bob mutually decide initial displacement $d=1$ for the  first QKD run and the displacement for subsequent runs can be calculated by $d = (p$ mod $m)$+1, where $p$ bits is the previous key length and the key length in the current QKD run is $m$ bits. Other methods can also be employed for choosing a random displacement using the secret key generated. The randomness in the choice of displacement ensures that there is no correlation between the position of the $m^{\rm{th}}$ check bit and the stego bit. It is to be noticed that to any third party, \textbf{MQS6} raises no suspicion and the protocol looks like an innocent QKD run. Additionally, the protocol is self-sufficient as it also generates the secret key needed for the next run. Despite this, MQS protocol has some weaknesses regarding which the two main points worth highlighting are:
\begin{description}
\item{\bfseries Direct communication:-} The party who wishes to share the stego bit prepares and sends the initial qubits. Also, the check bits are announced by the same party  in order to communicate the stego bit. Since the classical communication is in the same direction as that of transmission of qubits, we refer it to as "direct communication" here.
\item{\bfseries Embedding rate:-} Since MQS protocol allows only one stego bit to be embedded in the QKD protocol, the protocol is not efficient. Alice may embed her stego bit in a key of shorter length to increase efficiency.
\end{description}
In the next section, we will discuss how, together with a high embedding rate, direct communication proves fatal to the secrecy of the hidden channel. 
\section{Steganalysis of MQS protocol\label{sec:steganalysis}}
 The meaning of steganalysis can be easily understood from the fact that the relation between steganography and steganalysis is analogous to the relation between cryptography and cryptanalysis. Thus,  steganalysis aims to prevent covert communication or steganography. It usually targets to detect the deviation from theoretical probability distribution of states without being detected, i.e., in the context of MQS protocol QKD efficiency must not decrease beyond a certain threshold. It differs from a standard attack on the QKD protocols as it only aims to detect the deviation from a uniform key while standard QKD attacks are applied in order to gain complete information about the key. If a deviation is detected in steganalysis, the eavesdropper can, then, use standard QKD attacks in order to get the stego bit with high probability. In steganalysis of MQS protocol, Qu et al. \cite{stega} have applied ambiguous quantum state discrimination (QSD) attack to analyze the variation of the difference between the probability distribution of detected states from the theoretical distribution with respect to the embedding rate, using two measures of coherence.

In ambiguous QSD attack \cite{Holevo2001,Helstrom1969Jun}, the eavesdropper Eve tries to discriminate between the random states from the ensemble $\{\rho_i, p_i \}^d_{i=1}$ where $\rho_i = \ket{\psi_i}\bra{\psi_i}$ are the transmitted states  with probabilities $p_i$ and identify the incoming state. For this purpose, in general, Eve can use positive operator valued measurement (POVM) to perform measurements. A POVM is a set of positive semi-definite matrices $\{M_{i}\}$ that satisfy the completeness \cite{pathak2013elements}. If the system is in the state $\rho_i$, the maximal success probability to identify $\{\rho_i, p_i \}$ is 
\begin{equation}
    P_{\rm{success}}^{\rm{opt}}(\{\rho_i, p_i \}) = \max_{{M_{i}}}\sum_{i=1} p_iTr(M_{i}\rho_i),
\end{equation} obtained by maximization over all POVMs. The corresponding minimal discrimination error probability (MDEP) is 
\begin{equation}
    P_{\rm{error}}^{\rm{opt}}(\{\rho_i, p_i \}) = 1-\max_{{M_{i}}}\sum_{i=1} p_iTr(M_{i}\rho_i).
\end{equation}
MDEP is maximal when the prior probabilities are equal and changes accordingly with variation in the probabilities $ p_i$.\\ 
For our discussion, it is convenient to define the embedding rate $E$ as the ratio of number of secret bits communicated to the total number of bits in a single QKD run (cf. \cite{stega} ): 
 \begin{equation}
     E = \frac{\text{Number of stego bits}}{\text{Total number of QKD
     bits}}.
     \end{equation}
The probability distribution of states $\{\ket{\psi_i}, p_{i} \}^4_{i=1}$in  MQS protocol given in terms of "$E$" is: 
\begin{equation}
  \{p_i\}= \left\{\frac{n-nE}{4n}+ \frac{nE}{2n}, \frac{n-nE}{4n} , \frac{n-nE}{4n}+ \frac{nE}{2n}, \frac{n-nE}{4n} \right\} = \left\{\frac{1+E}{4}, \frac{1-E}{4}, \frac{1+E}{4},\frac{1-E}{4}  \right\},    
 \end{equation}
 where $\{\ket{\psi_i}\} = \{\ket{0},\ket{1},\ket{+},\ket{-}\}$. The distribution of the classical bits `0' and `1' is $\{(1+E)/2, (1-E)/2\}$. Equation (4) shows that as the embedding rate in the protocol changes, the probability distribution of states shows a substantial amount of change which is also reflected in the MDEP (Equation (2)). When the embedding rate is low, the MDEP changes only slightly but decreases significantly with high embedding rates, making it easier to successfully identify the quantum state. Also, the difference in the detected distribution of the transmitted states from the initial probability distribution itself raises the suspicion of covert communication between Alice and Bob, compromising the secrecy of the steganographic protocol.
\section{Reverse communication quantum steganography\label{sec:new-protocol}}

Here, we propose that the above kind of steganalysis can be evaded by modifying the standard "direct communication" steganographic protocols to include "reverse communication", while the other key features of the protocols are retained, such that no suspicion arises. To the best of our knowledge, this is the first time that this simple fix to the above steganalytic attack has been put forth. Reverse communication steganography works on the principle that if Alice wants to communicate a stego bit, she asks Bob to start the QKD run (similarly when Bob wants to communicate the stego bit, he asks Alice to start the QKD run). The classical communication is in the opposite direction, hence the name. As an example consider the case in which Alice wishes to communicate the stego bit.  Then, when she receives the qubits from Bob, she announces the conclusive results with a slight variation: specifically, while announcing the positions of the conclusive results, she deliberately announces an inconclusive result as the last conclusive bit, where it has a pre-decided displacement $d$ (calculated by the previous QKD run) with the actual conclusive result. Another way is that Alice announces her conclusive results randomly with the condition that the $d^{\rm{th}}$ announcement is the stego bit. In this manner, since Bob is aware of the scheme, he can know the stego bit.

The requirement of security comes from the QKD protocol itself while that of secrecy against steganalysis is fulfilled pertaining to the fact that Bob holds no knowledge about the stego bit Alice wishes to send, hence he would send all states with equal probabilities. Therefore, using this "reverse communication" procedure, there is no deviation of the detected distribution from the theoretical probability distribution, rendering the steganalysis attack useless. In the next section, we develop such protocols explicitly. Initially, reverse communication quantum steganography for four state protocol is proposed. Further, it is generalized to the whole class of discrete modulation CV-QKD protocols using coherent states. It is also shown that the proposal can be extended to other protocols of the same class that employ different states, namely, photon added subtracted coherent states (PASCS).

\subsection{Discrete Modulation CV-QKD Protocols} In CV-QKD systems, quantum states with infinite dimensional Hilbert spaces are used, where the information can be encoded in the position and momentum quadratures. The difference in discrete variable (DV)- and CV-QKD lies in the fact that in the latter, real amplitudes are measured instead of discrete events. Instead of photon counting techniques, the protocols employ homodyne detection. There are two classes of CV-QKD protocols:
        \begin{itemize}
            \item \textbf{Discrete modulation CV-QKD protocols} - The encoding is discrete in nature since the states used are simply mapped to binary bits \cite{Hirano_2017}.
            \item \textbf{Gaussian modulation CV-QKD protocols} -  In such protocols \cite{Grosshans_2003,Grosshans_2002}, the key is encoded as the real values of a Gaussian distribution, from which a key can be distilled using classical post-processing. These protocols are efficient over short ranges and can be implemented easily.
            \end{itemize}
Discrete modulation CV-QKD protocols can be seen as a direct analogue of standard DV protocols. Along with being experimentally feasible \cite{Hirano_2017}, they also provide security \cite{Ghorai_2019,Kaur_2021,Lin_2019} under a number of attacks similar to their discrete counterparts. For the aforementioned reasons, here, we propose a steganographic protocol that employs the use of such protocols. Majority of discrete modulation CV-QKD protocols are built in such a way that the classical communication from Alice's end is minimized to a certain extent, often eliminated completely. This inherent property provides them with a certain advantage to integrate the reverse communication steganography without any major modifications. 
\subsection{Steganographic communication using coherent state protocols}
\subsubsection{E4 Protocol}
Originally, the four coherent-state (O4) protocol \cite{Namiki_2003,PhysRevA.74.032302} was introduced using phase encoded coherent states (Fig.\, 1a). In \cite{PhysRevA.74.032302}, a wide class of discrete modulation CV-QKD using coherent states was introduced, which generalized and improved the O4 protocol. These protocols exploit the symmetries of the phase space to improve the efficiency while keeping the security same as before.
\begin{figure}[h]
\centering
  \subfloat[]{\includegraphics[width=0.4\textwidth]{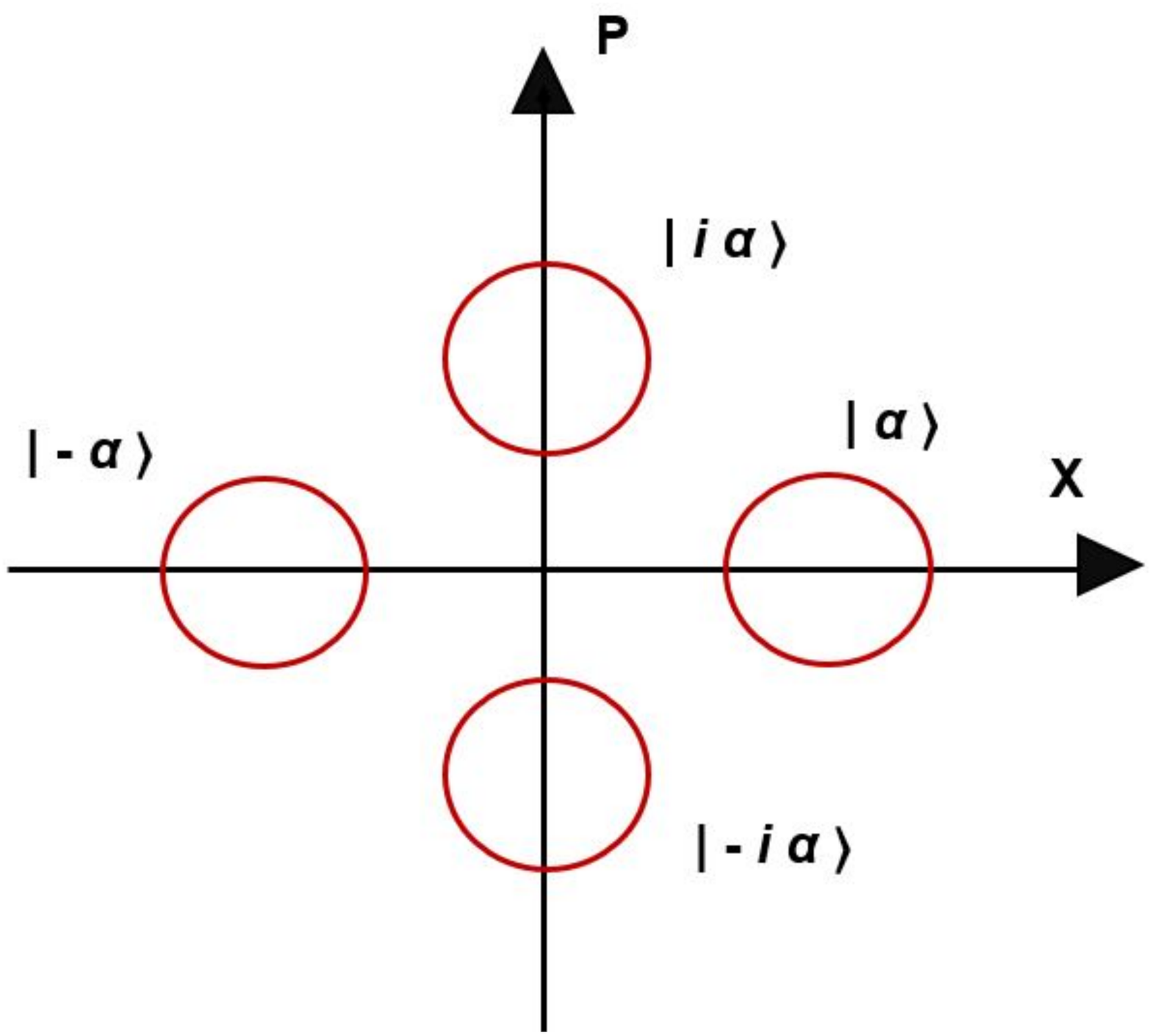}\label{fig:f1}}
  \hfill
 \subfloat[
  ]{\includegraphics[width=0.4\textwidth]{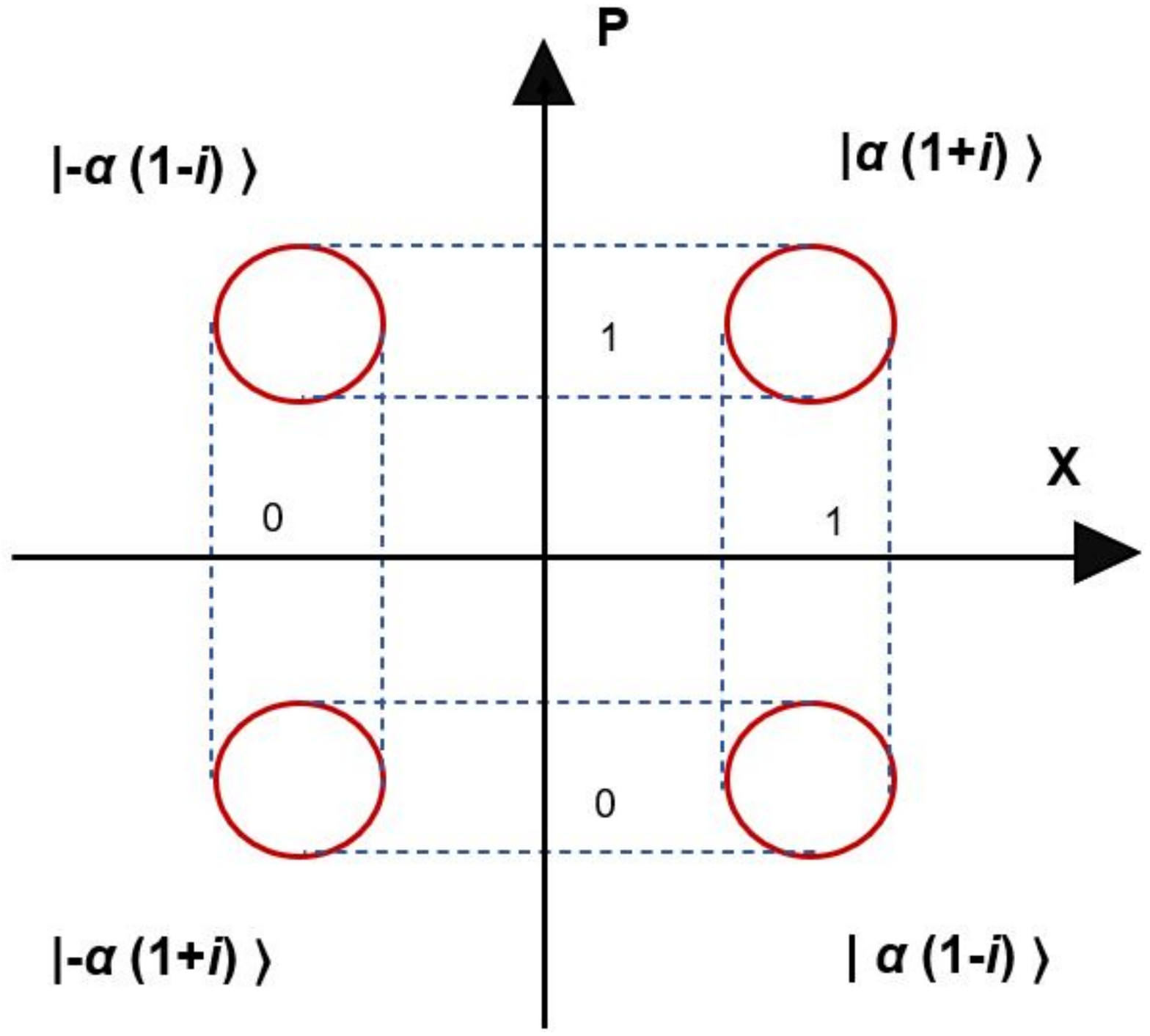}\label{fig:f2}}

  \caption{(Color online) Phase encoding protocols using coherent states: (a) The O4 protocol uses coherent states that have a symmetry about the origin in the phase space. (b) The E4 protocol with Alice's encoding corresponding to the quadratures.}
  \label{fig:O4E4}
\end{figure}
In order to understand this better, we review the "efficient" version (E4) of the four state protocol here:
\begin{enumerate}
    \item Alice sends randomly one of the coherent states $\ket{\alpha (1+i)}$, $\ket{\alpha (1-i)}$, $\ket{-\alpha (1+i)}$, $\ket{-\alpha (1-i)}$ with $\alpha >$ 0 (where $\mu=\lvert \alpha \rvert^{2}$ is the average number of photons) to Bob.
    \item Bob measures the position ($\hat{x}$) or momentum ($\hat{p}$) quadrature randomly.
    \item Bob informs Alice of his measurement quadrature. For the state $\ket{\alpha (1+i)}$, Bob's measurements yields a Gaussian distribution centered at $+\alpha$ in both the quadratures.\\ Similarly, for the state $\ket{\alpha (1-i)}$, Bob's measurements yields a Gaussian distribution centered at $+\alpha$ in the position and $-\alpha$ in the momentum quadrature.\\ For $\ket{-\alpha (1+i)}$, Bob's measurements yields a Gaussian distribution centered at $-\alpha$ in both the quadratures.\\ For $\ket{-\alpha (1-i)}$, Bob's measurements would yield a Gaussian distribution centered at $-\alpha$ in the position quadrature and centered at $+\alpha$ in the momentum quadrature. 
    \item Alice encodes her bit, corresponding to the measured quadrature, as depicted in Fig. \ref{fig:O4E4}(b).
    \item After the measurements, Bob simply announces the coordinates of the conclusive results, according to a post-selection threshold $x_0$. If the quadratures give the measured values above the threshold, the results are deemed as conclusive, otherwise inconclusive. This is necessary to mitigate the bit errors that may otherwise arise due to the strong overlap of quadrature distributions below this threshold.
\end{enumerate}
We define the efficiency $P_e$ of a protocol as the probability that the state is measured in the correct basis. Thus, $P_e$ of a protocol is a measure of the fraction of the measurement results that are not discarded during the execution of the protocol. Since no results are discarded in E4 protocol, its efficiency, $P_e = 1$. It is to be noticed that the initial states are prepared and sent by Alice and the conclusive results are communicated from the "reverse" direction, i.e., Bob's end. Hence, if Bob wishes to communicate a stego bit, he asks Alice to initialize the QKD run (and vice versa), following which Steps 1-4 are executed. For successful steganographic communication, appropriate changes are to be made in Step 5 of the above protocol in the following manner.\\
\\
\textbf{5$\rq$} Bob announces his conclusive results randomly with the condition that the $d^{\rm{th}}$ announcement is the stego bit.
\subsubsection{Generalization to $N$-state protocols}
The generalization of these protocols to $N$-state protocols has been explicitly shown in \cite{PhysRevA.74.032302}. The comparison for $P_e$ for $N = 3, 4, 6, 8$ has been shown in Table \ref{tab:abc}.
The general $N$-coherent state protocol works as follows:
\begin{enumerate}
    \item Alice sends $n$ coherent states to Bob for any $N$-state protocol. 
    \item Bob measures the position ($\hat{x}$) or momentum ($\hat{p}$) quadrature randomly.
    \item \textbf{Classical Communication:} Bob announces his measurement basis.
    \item \textbf{Classical Communication:} Alice confirms whether the measurement basis is correct or not.
    \item \textbf{Classical Communication:} Bob announces the conclusive results.
    \end{enumerate}
The above steps reveal the inherent symmetry and provide a general structure of the similar protocols, which in turn allows one to develop other such protocols for different number of states, with different values of $P_e$.
\begin{table}[h]
\centering
\begin{tabular}{|l | l | l| l| l| l| l|}
\hline
Protocol & $N$- state & O4 & E4 & Three-state & Six-state & Eight-state\\
\hline
Efficiency, $P_e$ & $\dfrac{2+N}{2+2N}$ & 1/2 & 1 & 2/3 & 2/3 & 3/4 \\
\hline
\end{tabular}
\caption{Comparison of efficiency for different $N$-state protocols.}
\label{tab:abc}
\end{table}
Although there is classical communication from Alice's end in Step 4, the information of the correct measurement basis cannot be used by the eavesdropper to gain additional information about the states, which is evident from the facts that the encoding subsets in any $N$-state protocol have common states and these protocols are secure against standard beam splitter attacks. Therefore, it is secure against standard QKD attacks and more importantly, does not interfere with successful steganographic communication. Assuming that Bob is the party communicating the stego bit, for achieving the same, Step 5 is modified slightly as follows.\\
\\
\textbf{5$^{\prime}$} Bob announces his conclusive results randomly with the condition that the $d^{\rm{th}}$ announcement is the stego bit.
\subsection{Extension to CV-B92 protocol}
 In \cite{Borelli_2015}, it was shown that discrete modulation CV-QKD is possible using single PASCS. The nonclassicality of PASCS can be exploited for improving QKD performance compared to coherent states.\\
A single PASCS is obtained by simply adding, then subtracting a photon from a coherent state. It is defined as \begin{equation}
    \ket{\psi (\alpha)} = N_\alpha^{-1/2} \hat{a}\hat{a}^{\dagger}\ket{\alpha},
\end{equation}
where $\ket{\alpha}$ is the initial coherent state with $\alpha >$ 0 and the mean photon number $\lvert \alpha \rvert^2$, $\hat{a}$ and $\hat{a}^{\dagger}$ are the corresponding annihilation and creation operators and $N_\alpha$ is the normalization constant which is defined as $N_\alpha =\lvert \alpha \rvert^4 + 3\lvert \alpha \rvert^2 + 1$. $\ket{\psi (\alpha)}$ can be written as a superposition of a coherent state (Gaussian component) and a photon added coherent state (non-Gaussian component), i.e., \begin{equation}
    \ket{\psi (\alpha)} \propto\ket{\alpha} + \alpha \hat{a}^{\dagger}\ket{\alpha} .
\end{equation}\\
This protocol, which we refer to as CV-BB84 protocol here simply replaces the coherent states $\{\ket{\alpha}, \ket{-\alpha}, \ket{i\alpha}, \ket{-i\alpha}\}$ in O4 protocol with corresponding single PASCS $\{\ket{\psi (\alpha)}, \ket{\psi (-\alpha)}, \ket{\psi (i \alpha)}, \ket{\psi (-i \alpha)}\}$, respectively. The $\ket{\psi (\alpha)}$($\ket{\psi (i \alpha)}$) represents bit `1' and $\ket{\psi (-\alpha)}$($\ket{\psi (-i \alpha)}$) represents bit `0'.\\
Recently, a protocol that uses PASCS states and can be seen as a CV counterpart of the B92 protocol was also proposed \cite{Srikara2020Oct}. It eliminates the need for post-protocol classical communication from Alice's side completely.
The protocol can be described as follows:
\begin{enumerate}
    \item Alice randomly sends $\ket{\psi(\alpha)}$, corresponding to `0' or $\ket{\psi(i\alpha)}$, corresponding to `1'.
    \item Bob measures the position ($\hat{x}$), corresponding to `0' or momentum ($\hat{p}$) quadrature, corresponding to `1' randomly.
    \item Bob encodes his bit as
    \begin{equation*}
    bit = \begin{cases}
               1               & \text{if} \; x < -x_0\\
               0               & \text{if} \; p < -x_0\\
            \rm{inconclusive} & \text{otherwise}
           \end{cases}
\end{equation*}
where $x_0$ is the post-selection threshold which implies that the result is conclusice if and only if the measured value of $x$ (or $p$) is less than $-x_0$. 
\item After the measurements, Bob simply announces the coordinates of the conclusive results. Since Bob's bit values is exactly anti-correlated to Alice's encoding, he flips his bits in order to obtain the common mutual secure key.
\end{enumerate}
The absence of direct communication makes perfect recipe for successful reverse steganographic communication. This can be achieved by the following modification in Step 4:\\
\\
\textbf{4$\rq$} Bob announces his conclusive results randomly with the condition that the $d^{\rm{th}}$ announcement is the stego bit.
\section{Conclusion\label{sec:conclusion}}
Motivated by the vulnerabilities of the MQS protocol  based on QKD, we discuss a general procedure to extend a QKD protocol into one for steganography in a way that eliminates this weakness. The basic, simple idea is that the party communicating the steganographic information is in the reverse direction of the party sending the initial quantum states. Further, the latter itself is camouflaged as if it is part of the classical reconciliation required for the QKD. This is demonstrated  through a number of example protocols. As CV-QKD has been experimentally demonstrated \cite{jouguet2013experimental, zhang2020long}, our proposed scheme is feasible with current quantum technology. Our work is an attempt  to explore the possibilities for steganography using QKD and we indicate a few new directions that it opens up for future exploration.

First is the question of the possibility of our proposal being adapted for protocols for secure direct communication \cite{yadav2014two, lucamarini2005secure}, which normally eliminate or minimize classical information being sent by Alice. Another is to explore other forms of reverse communication that may be used by Bob that can improve secrecy or classical efficiency. Lastly, our work is hoped to open up various possibilities for experimentalists. \color{black}

\section*{Acknowledgment} RJ, AG, RS and AP acknowledge the support from the QUEST scheme of Interdisciplinary Cyber Physical Systems (ICPS) program of the Department of Science and Technology (DST), India (Grant No.:
DST/ICPS/QuST/Theme-1/2019/14 (Q80)). KT acknowledges GA CR (project No. 18-22102S) and support from
ERDF/ESF project `Nanotechnologies for Future' (CZ.02.1.01/0.0/0.0/16 019/0000754). RS also acknowledges the support of DST, India, Grant No. MTR/2019/001516.

\bibliographystyle{unsrt}
\bibliography{akhroh}

\end{document}